\documentclass[]{spie}  
\usepackage[]{graphicx}
\usepackage{url}

\title{CRUSH: fast and scalable data reduction for imaging arrays}

\author{Attila Kov\'{a}cs
\skiplinehalf
MPIfR, Auf dem H\"{u}gel 69, 53121 Bonn, Germany
}

\authorinfo{Further author information: E-mail: akovacs@mpifr-bonn.mpg.de, Telephone: +49 228 525 196}

\begin{document}


  \maketitle

\begin{abstract}
{\em CRUSH} is an approach to data analysis under noise interference, developed specifically for submillimeter imaging arrays. The method uses an iterated sequence of statistical estimators to separate source and noise signals. Its filtering properties are well-characterized and easily adjusted to preference. Implementations are well-suited for parallel processing and its computing requirements scale linearly with data size -- rendering it an attractive approach for reducing the data volumes from future large arrays.
\end{abstract}


\keywords{data reduction, imaging, bolometer arrays, submillimeter, correlated noise, sky noise}

\section{INTRODUCTION}
\label{sec:intro}  

The arrival of increasingly larger imaging bolometer arrays, operated in total-power scanning observing modes, pose new challenges for data reduction. Viable approaches may have to cope with an atmospheric foreground (in case of ground-based instruments), whose brightness and typical variance dwarf astronomical signals, and also with instrumental imperfections, such as temperature fluctuations of the detector plate, $1/f$ type noise, and various electronic interference along the signal paths.

Moreover, the typical data rates of next-generation bolometer instruments, in the range 20--200\,Hz, mean that the kilo-pixel-scale arrays of the future (E.g.~SCUBA-2\cite{scuba2} or ArTeMiS\cite{artemis}) will churn out data in the hundreds-of-megabytes to tens-of-gigabytes range per hour. Such volumes of data are not easily handled. Brute-force matrix approaches\cite{matrix} may become too slow for practicality (as matrix multiplications scale $\mathcal{O}(N^2)$ and inversions can scale up to $\mathcal{O}(N^3)$ with the data size $N$). Faster methods would be welcome, especially if their computational costs increase linearly with data volume. In addition, data reduction approaches suited for distributed computing, would be able to reduce full data sets organically and in a self-consistent manner (up to perhaps the hundreds of hours of integrations needed for large-area and/or deep-field surveys) on a cluster of networked computers.

One such approach that meets both requirements of linear computation time, and distributability over computing nodes, is {\em CRUSH}. It's data reduction concept already powers some of today's large bolometers\cite{sharc, laboca}, and has proven itself capable of producing images with unsurpassed quality. Thus, the aim of this paper is to introduce this approach and give a practical description of its structure and algorithms. For the detail hungry, a more thorough and mathematically motivated description is offered by Ref.~\citenum{thesis}.

\subsection{Origins and Future}

{\em CRUSH}\cite{thesis} (Comprehensive Reduction Utility for SHARC-2) began as the designated data-reduction suite for the 12$\times$32 pixel SHARC-2 (Submillimeter High-Angular Resolution Camera 2)\cite{sharc}. SHARC-2 was one of the first bolometer arrays to break with the traditional position-switched differencing (chopping) observing mode, and embrace scanning in total power instead. The new mode, however, required a new data reduction approach also, to effectively separate the faint astronomical signals from a bright and variable atmospheric foreground, and to overcome adverse noise signals originating from within the instrument (e.g. cold-plate temperature fluctuations, bias drifts or electronic pickup, $1/f$ noise). Since then {\em CRUSH} has proved itself capable of producing high-quality images that surpass chopped images of the past both in fidelity and in clarity.

The original {\em CRUSH} software package\footnote{See \url{www.submm.caltech.edu/~sharc/crush/}.} is not the only software that uses its data reduction philosophy. Other packages, like {\tt sharcsolve} developed by C.D.~Dowell ({\em Caltech)} in parallel with {\em CRUSH} as an independent implementation and cross-checking platform for SHARC-2 data reduction, and {\em BoA}\cite{boa} under development at the Max-Planck-Institute for Radio Astronomy (MPIfR) mainly for the {\em Atacama Pathfinder Experiment}\cite{apex} (APEX) bolometers, follow the same reduction approach (even if BoA does this in an interactive form vs.~the preconfigured pipeline implementations of the others). {\em MiniCRUSH} is yet another streamlined alternative implementation by the author for the APEX arrays.

A more versatile and user-friendly version of {\em CRUSH}, capable of supporting a wider variety of instruments, and distributed computing, is currently under development by the author, and will be released within the year. It aims to extend the usefulness of the approach beyond imaging bolometers and apply it to various other instruments operating under adverse noise conditions, esp. when used in scanning (e.g.~in frequency) observing modes.

\section{BACKGROUND}

Total-power detector time signals are a complex mix of superimposed components signals, which are produced by different physical processes, at the various stages of the signal path. The digitized time-stream of a typical bolometer carries not only astronomically interesting signals, but also white noise (e.g.~Johnson noise and photon shot-noise), and interference from a bright and variable atmosphere, temperature fluctuations on the detector plate, RF-pickup, microphonic noise from moving cables (for semiconductor bolometers), $1/f$ noise, signals that originate in various stages of the electronics (resonances, inductive 50/60\,Hz pickup), etc. The challenge of data reduction is thus to separate out the astronomical signals from this rich mixture of masking noise.

The effectiveness of the separation, measured by the accuracy to which the ``true'' astronomical source can be reconstructed, depends heavily on the observing mode used to obtain the data. The observing mode is almost single-handedly responsible for arranging the source signals in such a way that they are distinguishable from the noise interference\cite{scanning}, regardless of what data reduction approach is followed. It provides an absolute limit to how well any analysis can perform. The best {\em CRUSH}, or other methods can do, is reaching that limit\cite{thesis}.

\subsection{A Pipeline Approach}

The basic concept behind {\em CRUSH}, is to focus on one signal component at a time -- estimating it, and subsequently removing from the time-stream -- then proceeding step-by-step onto the other components in a pipeline, peeling away interfering signals one at a time until only the astronomical source and white noise remains (or alternatively, until the source signal becomes sufficiently predominant in the time stream such that the remaining noise is no longer considered bothersome). This is in contrast to matrix methods, which perform the separation in a single go.

This makes {\em CRUSH} seem inefficient at first glance. However, there are a number of practical advantages to the step-by-step approach over the single-go methods. One of these is computational: the cost of {\em CRUSH} reduction grows linearly with data volume, whereas matrix inversions can require up to $\mathcal{O}(N^3)$ operations (compression of large data sets\cite{compression} can reduce this hungry requirement, albeit never to linear cost). There are other, more subtle advantages also, which are highlighted throughout this paper.

\begin{figure}
\label{fig:pipeline}
\includegraphics[height=0.76\textheight]{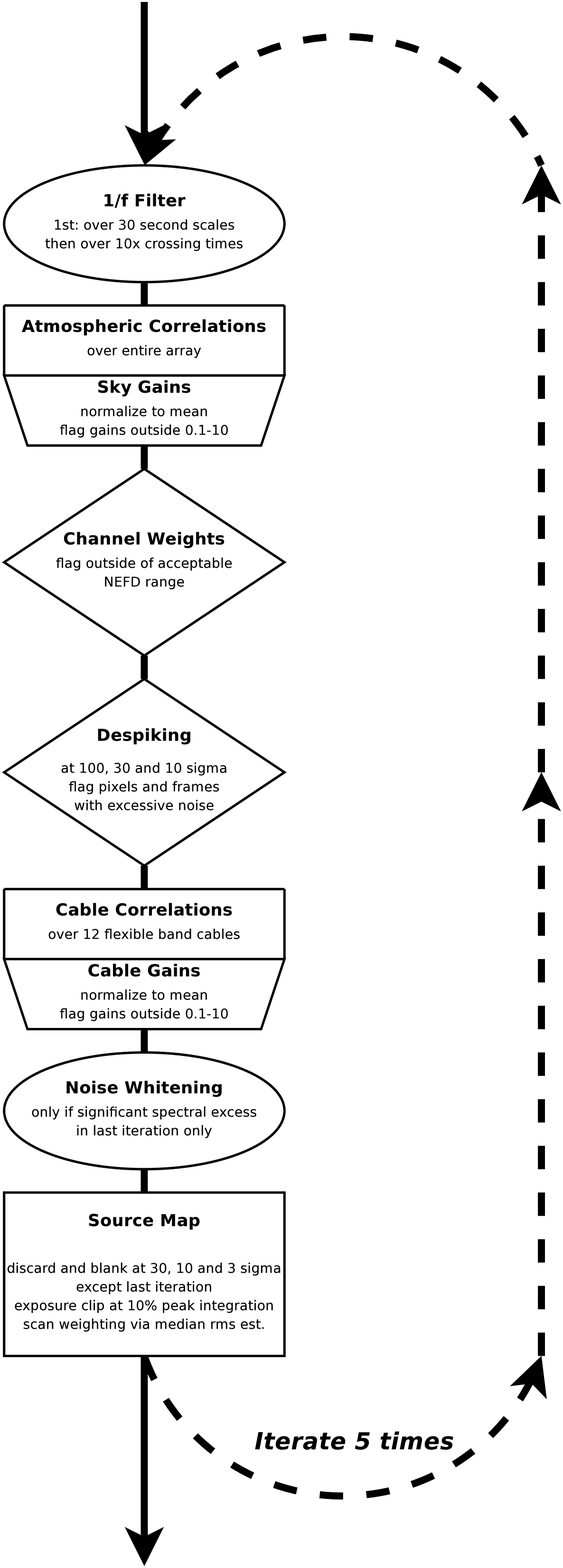}
\hfill
\includegraphics[height=0.76\textheight]{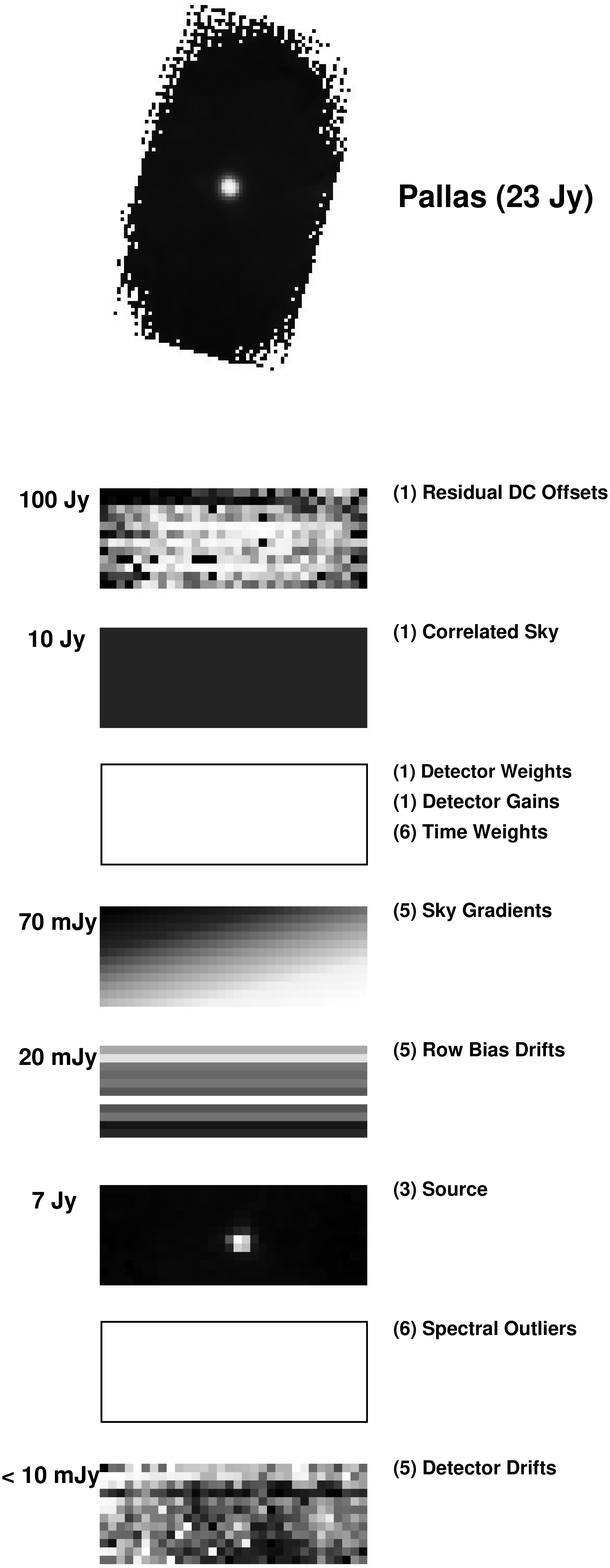}
\\
\\
\caption{Illustrations of typical {\em CRUSH} pipelines. The diagram on the left is a schematic of the default pipeline used for LABOCA reductions (v.~1.01). The various types of reduction steps are distinguished by shapes. Boxes represent the estimation of signal components, often followed by the estimation of the corresponding gains (trapezoids). Filtering steps are shown as ellipses, while data conditioning (weighting and flagging) are shown as diamonds. Five iterations of the pipeline are typically sufficient to reach acceptable convergence. Each reduction step can be activated, disabled or modified between iterations. For example, despiking starts at 100\,$\sigma$ level at first, and is tightened to 30, then 10\,$\sigma$ levels in consecutive rounds. The right side shows what modeled signals look like across the detector array in the default SHARC-2 pipeline (v.~1.52), at a selected frame of a scan on the asteroid Pallas. This pipeline was iterated 6 times. The starting iteration for each reduction step is indicated in parentheses. Typical signal levels are also shown. The last model, labeled as 'detector drifts' acts a time-domain $1/f$ filter that is estimated similarly to the other models. Gradients are estimated as correlated noise, with gain in direction $x$ explicitly chosen s.t.~$g_i \propto (x_i - \left< x \right>)$ for a pixel $i$ at projected position $x_i$.
}
%
\end{figure}

\subsection{Coping with the Adversities of Noise}

The interfering noise signals can be stationary or transient, correlated or individual to detector channels. Accordingly, it varies how they are best dealt with.

Correlated noise signals can be removed after estimating the correlated components from the time-stream. In essence, this is what matrix methods are designed to do for all correlated signals at once while also producing a source map in a single step. {\em CRUSH} estimates correlated components one by one, and makes the source map in a separate step. Yet the two approaches are mathematically equivalent, since the correlated components (e.g.~atmospheric fluctuation seen by the whole array, or electronic noise in one group of detector channels) are uncorrelated to one-another owing to their independent physical origins (or else these can be appropriately orthogonalized), and they should also be uncorrelated to the source signals, provided the scanning strategy was wisely selected\cite{scanning}.

Stationary noise processes produce well-defined spectral signatures, such as a $1/f$ type spectrum or lines. These can be optimally filtered, e.g.~by Wiener-filters or by noise-whitening\cite{thesis}. $1/f$ type noise can also be rejected in the time-streams using moving boxcar (or specifically shaped) windows\cite{thesis}, thus bypassing the need for costly Fourier transforms to and from spectral domains.

Transient noise, by nature, affects only certain areas of the time-stream. These data can be flagged (e.g.~spikes), or downweighted (e.g.~in case of temporarily elevated noise levels).

Matrix methods are not well adapted for incorporating either spectral filtering, weight estimation or flagging within them. Instead, these steps must be performed before or after the inversion. Therefore, practical matrix methods are not truly one-go.

In contrast, the step-by-step nature of {\em CRUSH} allows the ``interspreading'' of the extra filtering, weighting and flagging steps anywhere between the signal modeling steps. Figure~\ref{fig:pipeline} shows a flow-chart representation of the default pipeline used for LABOCA\cite{laboca} data reduction. The greater freedom to insert data conditioning steps (weighting, flagging, and filtering) between the various signal models has clear advantages. For example, after the removal of the strong correlated signals produced by the variable atmosphere, the bolometer time-streams are usually clean enough to allow weight estimation from noise and attempt a round of despiking. With the improved weights and a glitch-free time stream, the subsequent estimation of other signal components becomes instantly more reliable.

\subsection{Data Structure}

Before the discussion can focus on the details and merits of {\em CRUSH}, it is necessary to establish the basic concepts of the data structure it operates with. In an array instrument data arrive from multiple {\em channels}, which are digitized at (regular) intervals. The set of data consisting of individual {\em samples} from each channel for a given time interval constitutes a {\em frame}. A {\em scan} is a set of frames, for which both the instrument and the data taking environment can be considered stable (e.g.~gains and noise properties are essentially identical, the line-of-sight optical depth is approximately constant etc.). For typical submillimeter instruments this requirement is usually satisfied by single integrations. Were this not the case, integrations can be split or united until one arrives at chunks of data we can call a {\em scan}. The reduction steps of {\em CRUSH} operate on such {\em scans}, with the exception of the source model that may be derived from entire data sets.

\subsection{Signals and Notation}

The measured bolometer time-stream $X_c$ of channel $c$, is the composite of the source signals (i.e.~the source flux distribution $S$ mapped into the time-stream via the observing pattern $\mathcal{M}$ and with gains ${\bf G}$), the various correlated signal components ${\bf C}_1, ... {\bf C}_k$, which are present in the time-stream with gains $g_{1,c}, ... g_{{\rm k},c}$ respectively, and white (typically Gaussian) noise ${\bf n}$ as,

\begin{equation}
X_{ct} = G_{ct} \mathcal{M}_{ct}^{xy} S_{xy} + g_{1,c} C_{1,t} + ... + g_{{\rm k},c} C_{{\rm k},t} + n_{ct}.
\label{eq:timestream}
\end{equation}

The data reduction attempts to recover the source flux distribution ${\bf S}$ and the various correlated components ${\bf C}_{\rm i}$ from the measured data ${\bf X}$. If this is done based on sound mathematics, the resulting models $\hat{\bf S}$ of the source and $\hat{\bf C}_{\rm i}$ of the correlated signals should be unbiased estimates of the underlying physical signals, i.e.~$\hat{\bf S} \rightarrow {\bf S}$ and $\hat{\bf C}_{\rm i} \rightarrow {\bf C}_{\rm i}$. At times, it may become necessary to estimate the gains (${\bf G}$ and ${\bf g}_{\rm i}$) also, when these are unknown, or undetermined to sufficient accuracy beforehand.

Following the removal of one or more estimated components $\hat{\bf C}_{\rm i}$ or  $\hat{\bf S}$ with the best-guess values of gains $\hat{\bf g}_{\rm i}$ and $\hat{\bf G}$ (the hats indicating throughout that these are not the exact underlying components and true gains but rather our estimates for them) from the time stream, one is left with a residual signal ${\bf R}$, which is (in vector notation)

\begin{equation}
{\bf R} = {\bf X} - \hat{\bf G} \bullet (\mathcal{M} \cdot \hat{\bf S}) - \hat{\bf g}_1 \hat{\bf C}_1^\top - ... - \hat{\bf g}_{\rm k} \hat{\bf C}_{\rm k}^\top.
\label{eq:residual}
\end{equation}

Comparing the above to Eq.~\ref{eq:timestream}, it is clear that when the understanding of the signal structure is sufficiently complete and the estimates for source, gains and correlated noise are representative of the underlying signals, then ${\bf R} \rightarrow {\bf n}$, i.e., the residuals become essentially white noise.

\section{INSIDE {\em CRUSH}}

Now it is time to have a closer look at the typical reduction steps, which make up the {\em CRUSH} pipeline. This section discusses the estimation of correlated signal components, noise weights, and the identifying and flagging of bad data. The estimation of the astronomical source is deferred to the following section, owing to the pivotal importance of that step -- the very raison d'\^{e}tre of the reduction, -- as well as the various subtleties that accompany it.

\subsection{Correlated Noise Components}
\label{sec:correlated}

As already mentioned the heart of the {\em CRUSH} approach lies in the way it deals with the correlated and source signals -- step-by-step rather than at once. This is the main distinguishing point from matrix methods. All other steps, like filtering, weight estimation, and data flagging are common to all approaches.

Let us then focus on a single, correlated component (here just ${\bf C}$ without the distinguishing index $i$), or rather what is still left of it ($\Delta {\bf C} = {\bf C} - \hat{\bf C}$) in the residual time stream after possible prior steps of estimation and removal:

\begin{equation}
R_{ct} = ... + g_c\, \Delta C_t + ...
\end{equation}

For the moment forget everything else (which is anyway but noise from this perspective). Then consider the $\chi^2$, defined in the usual way as

\begin{equation}
\chi^2  = \sum_{t,c} \frac {R_{ct} - \hat{g}_c \Delta \hat{C}_t} { \sigma^2_{ct}},
\label{eq:chisquared}
\end{equation}

where $\sigma$ is the underlying rms (white) noise level for each sample. Notice that with the use of proper noise weights ($w = \sigma^{-2}$), $\chi^2$ can be rewritten as $\chi^2 = \sum w_{ct} (X_{ct} - \hat{g}_c \Delta\hat{C}_t)$. Thus, the $\chi^2$ minimizing condition $\partial \chi^2 / \partial (\Delta \hat{C}_t) = 0$, yields the maximum-likelihood estimate for $\Delta C_t$ as,

\begin{equation}
\Delta\hat{C}_t = \frac{\sum_{c} w_{ct} \hat{g}_c R_{ct}} { \sum_{c} w_{ct} \hat{g}_c^2 }.
\label{eq:estimator}
\end{equation}

The uncertainty $\sigma(\hat{C}_t)$ of this estimate can also be expressed considering the change in $\hat{C}_t$ required for increasing $\chi^2$ by 1:

\begin{equation}
\sigma^2(\hat{C}_t) = \frac{ 1} { \sum_{c} w_{ct} \hat{g}_c^2 }.
\label{eq:uncertainty}
\end{equation}

Clearly, the summations required for the calculation need to be performed only on the subset of channels that are sensitive to this particular correlated component (i.e.~for which $g_c \neq 0$).

\subsubsection{Time-Resolution and Filtering of Correlated Signal Models}
\label{sec:model-filtering}

Often, the correlated noise interference can be isolated into specific spectral components or regions (such as the low frequency bins in case of $1/f$ type noise or well-defined frequency intervals of resonances). Accordingly, their models can be restricted to those spectral areas by appropriate filters. The spectral filtering of the correlated signal models should be performed after the unbiased estimation of these (Eq.~\ref{eq:estimator}) and before the removal from the time-streams. For $1/f$ type noise signals, the discarding of unneeded high-frequency components from the models can be effectively achieved by extending the summations in Eqs.~\ref{eq:estimator} and \ref{eq:uncertainty} to regular time-windows ($T \sim 1/2f_{\rm knee}$)\cite{thesis}, i.e.~by lowering the time-resolution of the models.

Explicit spectral filtering of signals will reduce the effective number of parameters derived for them by a factor of $(1-\Phi)$, where $\Phi = \left<|\phi_f|^2\right>_f$ given the spectral filter coefficients $\phi_f$ (cf.~Parseval's theorem). This will be important to keep in mind when estimating noise weights (Sec.~\ref{sec:weighting}). 

In contrast to {\em CRUSH}, generic spectral filtering of correlated signal models is not easily accomodated within a matrix inversion step, and therefore has to take place separately once the set of signal solutions is obtained. As a result the filtering of some models will not benefit the solutions of others (esp.~the source) in the way the sequential nature of {\em CRUSH} allows. Matrix inversion methods thus offer no meaningful way of improving the quality of reduction through filtering of the correlated signal estimates.

\subsubsection{Gain Estimation}

Once an estimate of the correlated component in question is calculated, the successful total (or at least sufficient) removal of this correlated signal component from the time-stream depends critically on the accurate knowledge of the corresponding gains $g_c$, with which $C_t$ are mapped into the individual bolometer time-streams. Any the error $\Delta g_c = g_c - \hat{g}_c$ in the gain estimate will leave behind remnants of $\Delta g_c \hat{C}_t$ from the imperfect removal of $\hat{\bf C}$. Most ground-based bolometer cameras operate under a correlated atmospheric foreground whose variations can trump the faint astronomical signals up to $10^5$ times in magnitude. This means that knowledge of (relative) gains to $\ge$5 significant figures may be required before the faint source becomes at all visible\footnote{This also partly because the typical $1/f^2$ spectrum of sky noise means that it does not integrate down in time.}. One hardly ever knows the relative detector gains quite so accurately a priori. Certainly not for semiconductor bolometers, whose detector gains exhibit strong dependence on optical loading power.

Fortunately, the required gains can be estimated from the data itself, analogously to the correlated component. The corresponding $\chi^2$ minimizing condition $\partial \chi^2 / \partial (\Delta \hat{g_c}) = 0$ yields:

\begin{equation}
\Delta\hat{g}_c = \frac{\sum_{t} w_{ct} R_{ct} \hat{C}_t} { \sum_{t} w_{ct} \hat{C}_t^2 },
~~~{\rm and }~~~
\sigma^2(\hat{g}_c) = \frac{ 1} { \sum_{t} w_{ct} \hat{C}_t^2 }
\label{eq:gains}
\end{equation}

for the incremental maximum-likelihood gain estimate and its uncertainty respectively. Following the estimation, the hereby modeled $\Delta \hat{g}_c \hat{C}_t$ products are duly cleaned from the residual time-streams.

While gain estimation fits seamlessly within the {\em CRUSH} pipeline, usually in immediate succession to the estimation of the corresponding correlated component, it does not find an easy place in matrix methods, which require gain solutions to be estimated between iterated matrix inversions steps, alongside the other ``outcasts'' of weighting, flagging, and filtering.

\subsubsection{Gain Normalization}

One subtlety of gain fitting is that it opens up a degeneracy between gains and the corresponding correlated components. Multiplying one and dividing the other with the same constant value leaves the product $\hat{g}_c \hat{C}_t$ unchanged. Since it is the product that features in the time-stream, all those solutions are equally good from a mathematical point of view. This is not a problem unless one assigns physical meaning to the gain values or the correlated signals.

 Gains derived for one correlated component may be useful elsewhere (E.g.~the sky-gains resulting from the correlated atmospheric variations can be a factor in source gains), and the correlated atmospheric noise can be a source of improved line-of-sight opacity estimates (see Section~\ref{sec:opacities}).

 The degeneracy is broken is the gains are normalized. Typical gain normalizations may fix the average gain values (weighted: $\sum_c w_c \hat{g}_c / \sum_c w_c$, or arithmetic: $\sum_c \hat{g}_c / N$), or the gain variances (e.g.~$\sum_c w_c \hat{g}_c^2 / \sum_c w_c$), to unity or some predetermined ``absolute'' gain value.

\subsubsection{Gain Flagging}

It is often practical to use the normalized gain information when trying to spot and flag detector channels that behave oddly. Either because these exhibit too little response to the correlated component (meaning these are ``blind'') or because they respond too much (e.g.~going off the charts).

\subsection{Weight Estimation}
\label{sec:weighting}

Weighting is based on the simple idea of assigning proper noise weights ($w = 1 / \sigma^2$) to data. In principle each datum can have an independent weight value $w_{ct}$. However, as the underlying noise $\sigma_{ct}$ is usually separable into a stationary channel noise $\sigma_c$ and a time-variant noise component affecting all channels $\sigma_t$, the separation can be carried on to weights; i.e., $w_{ct} = w_c \cdot w_t$. Provided that the underlying noise variance is estimated as $\sigma^2 = \sum R^2 / (N-P)$ for a data set with $P$ lost degrees-of-freedom (i.e.~parameters), we can calculate the weight components according to:

\begin{equation}
\hat{w}_c =  (N_t - P_c) \frac { \sum_t w_t } { \sum_t w_t R_{ct}^2 },
~~~{\rm and}~~~
\hat{w}_t = (N_c - P_t) \frac { \sum_c w_c } { \sum_c w_c R_{ct}^2 },
\end{equation}

where $P_c$ and $P_t$ are the effective total number of parameters derived (i.e.~the lost degrees of freedom) from time time-stream of channel $c$ or from frame $t$. This reflects the fact that by crunching pure noise through the reduction, its rms level will be artificially decreased from its underlying level due to partial 'modeling' as correlated components etc.

As for calculating the exact number of parameters ($P_c$ and $P_t$) derived from the data of a given channel or frame, consider Eq.~\ref{eq:estimator}, and note how much each data point contributes to the given estimate. Since correlated signals, gains (Eq.~\ref{eq:gains}) and source map (Eq.~\ref{eq:source}) are derived in similar manner, the same extends to all parameters of the reduction. Consider then an estimate of some parameter $\hat{A}_{\rm i} = \sum w g R / \sum { w g^2}$. Each point in the summation contributes a part $p_{{\rm i},ct} = w_{ct} g_{ct}^2 \sigma_{\rm i}^2$ to the estimate (where $\sigma^2_{\rm i} = 1 / \sum w g^2$). Hence, the lost degrees of freedom due to the estimation of all parameters $\hat{A}_{\rm i}$ can be calculated as $P_c = \sum_{\rm i} (1-\Phi_{\rm i}) \sum_t p_{{\rm i},ct}$ for channel $c$ and $P_t = \sum_{\rm i} (1-\Phi_{\rm i}) \sum_c p_{{\rm i},ct}$ for frame $t$, after accounting for the effective reduction of parameters (by $1-\Phi_{\rm i}$) due to possible filtering of raw models (Sec.~\ref{sec:model-filtering}).

The critical importance of such careful accounting is often underappreciated. However, failure to keep proper track of the lost degrees-of-freedom will result in unfair weight estimates. While initial deviations from fair weights may be small, these tend to diverge exponentially with successive iterations, resulting in unstable solutions with runaway weight estimates in iterated reduction schemes, such as is necessary for bolometer data (see Sec.~\ref{sec:discussion}).

Another point to look out for is to break the degeneracy, under multiplication by a scalar, between the two weight components $w_c$ and $w_t$. The practical approach is to fix the normalization of time weights s.t.~$\left< w_t \right> = 1$.

Weights can also be estimated using alternative methods, such as using median square deviation, instead of the maximum-likelihood estimation presented above. For medians a useful approximate relation to remember is med$(x^2) \approx  0.454937 \sigma_x^2$ (see Ref.~\citenum{thesis}).

Note also, that like gain estimation, weighting must take place outside the matrix inversion, when matrix methods are used for separating signals. Thus weighting adds to the growing list of unavoidable reduction steps that render matrix approaches into a multi-step process.

\subsection{Flagging Bad Data}

Identifying and flagging bad data can be critical in getting optimal results. The detector time-streams may contain glitches, such as produced by small electronic discharges, cosmic rays, mechanical jolts etc. At times detectors, which are normally well-behaved, can become finicky and troublesome. Time-streams may also have spurious spectral resonances that remain untackled in the reduction. Unless one keeps an eye out for such data defects and removes these troublesome data points from further analysis, the reduction quality will be compromised. The number, type and details of algorithms looking for bad data are limited only by the creativity of mind. Therefore, instead of giving recipes, a few examples of what sort of troubles one might look for is presented here.

The case of the very occasional electronic glitch is easily tackled by the simplest despiking methods. These typically look for absolute deviations in the data that are well in excess of what could be justified under a normal-looking noise distribution.

At times the problematic data resides in more than an occasional single data point. Transition-Edge Sensor (TES) bolometers can contain discontinuities resulting from the branching of the SQUID readout with changing flux. Also, the APEX bolometers often see transient glitches in the time-stream that can span up to a few seconds in duration.

For problems in the time-stream spectra, such as unmodeled and unfiltered resonance peaks, the above methods, which seek glitches and wider features, can be adapted from the time domain to spectral domains.

The weights and gains derived during the reduction (see above Sections), can serve as useful diagnostics. A good practice can be to discard any channel (or frame) that has unreasonable weights and/or gains. Clearly, channels with low weights and/or gains are insensitive and contribute little or nothing to all estimates (including the source model). On the flopside, gains and weights that are unrealistically higher than the array average, are unlikely to be physical and could signal some serious malfunction of that channel. Channels and frames that are left with no degrees-of-freedom in should also be flagged, as these no longer contain useful information.

Finally, some practical notes on flagging. In a iterated scheme each round of flagging should revise prior flags, allowing data previously judged to be problematic to become unflagged and re-enter the analysis, provided these now appear well-behaved. It is also advisable to keep different types of flags (e.g. spikes, jumps, gain flags, degrees-of-freedom flags etc.) separately accounted for. Lastly, it is worth mentioning that flagging constitutes yet another reduction step, which must be performed outside of the inversion, when a matrix approach is used.

\subsection{Alternative Statistical Estimators}
\label{sec:estimators}

Thus far, the models for correlated signals, gains and weights were derived using the maximum-likelihood estimates that resulted from $\chi^2$-minimizing conditions. However, {\em CRUSH} leaves the door open for other statistical estimates also. For example, one may replace the weighted means of the maximum-likelihood estimates with robust measures (like medians or weighted medians\cite{thesis} or trimean). These have the advantage that they can remain unbiased by the presence of strong source signals or spikes (although one should note that faint sources below the time-stream noise level will bias such estimates also!). The drawback of median-like estimates is that their computation requires sorting, which is an $\mathcal{O}(N \log N)$ process with element number $N$ versus the strict linearity of maximum-likelihood estimates. Maximum-entropy estimates can be derived as a correction to the $\chi^2$-minimizing ones\cite{thesis}.

Whichever statistical estimation method is used, the formulae derived for the uncertainties and the lost degrees-of-freedom in maximum-likelihood estimates, will hold throughout. This is because, under symmetric white noise (e.g.~Gaussian noise), all estimates provide measures of the same underlying quantity, which is the center of the noise distribution. As the uncertainties and lost degrees of freedom depend only on the noise properties, and not on the presence of other signals, these remain fully valid.

Matrix methods are unsuitable for using median-like estimates, which require sorting. On the other hand maximum-entropy corrections can be applied in a separate step outside the matrix inversion. The flexibility of using a statistical estimator of choice is an attractive feature of the {\em CRUSH} approach.

\section{SOURCE MODEL (MAP-MAKING)}
\label{sec:source}

Implementations of {\em CRUSH} typically use the nearest-pixel mapping algorithm, mainly because it is the fastest (and linear with data volume) and most direct way of producing maps. It also proved sufficient in arriving at high-fidelity source maps. The algorithm uniquely associates each time-stream data point $X_{ct}$ to a map pixel $S_{xy}$, in which the data point ``deposits'' all the flux it carries. The maximum-likelihood incremental source model is:

\begin{equation}
\Delta \hat{S}_{xy} = \frac { \sum_{ct} \delta_{ct}^{xy} w_{ct} \hat{G}_{ct} R_{ct} }
{ \sum_{ct} \delta_{ct}^{xy} w_{ct} \hat{G}_{ct}^2 }.
\label{eq:source}
\end{equation}

 Here the unique association of time-stream data $X_{ct}$ to a map pixel $S_{xy}$ is captured by the Kronecker-delta $\delta_{xy}^{ct}$, which serves as the effective mapping function. In practice, one solves for $\hat{\bf S}$ in a single pass over all data, accumulating the numerator and denominator separately for each map pixel indexed as $\{ xy \}$. The renormalizing division is performed as a final step. The separate keeping of the gain-corrected weight-sum in the denominator has further use in estimating the uncertainty of the updated map flux values, i.e.:

\begin{equation}
\sigma^2(\hat{S}_{xy}) = \frac { 1 } { \sum_{ct} \delta_{ct}^{xy} w_{ct} \hat{G}_{ct}^2 }.
\label{eq:source-rms}
\end{equation}

There is no reason why other, more complex, mapping algorithms\cite{tegmark} could not be used in performing this step. However, the simplicity of the nearest-pixel mapping is attractive from a computational point of view, and sufficient for producing high-fidelity models of submillimeter sources, provided a fine enough grid of pixelization is chosen. Typically, 5 or more pixels/beam (FWHM) give highly accurate results, but as few as 3 pixels per beam can be sufficient for source maps of reasonable quality. (The default SHARC-2 reduction uses $\sim$6 pixel/beam, whereas the APEX bolometer reductions settle at 3--5 pixels/beam.)

The gain ${\bf G}$ used for mapping time-stream signals into the source map, is a composite of the relative atmospheric noise gains $g_c$ (derived from the correlated atmospheric noise according to Eq.~\ref{eq:gains}), a relative main-beam efficiency $\eta_c$ for each pixel, a atmospheric transmission coefficient $T_t$ (whose time variation may be estimated, see below), a (potetially loading dependent) calibration factor $Q$ which converts the time-stream counts or voltages to physically meaningful flux units, and filtering corrections (see Sec.~\ref{sec:filtering}) by $(1-\phi_c)$. I.e., $G_{ct} = Q \eta_c g_c T_t (1-\phi_c)$.

Because the maps are the ultimate goal of the data reduction, they can undergo post-processing steps, which shape them further to preference. The maximum-likelihood maps of Eqs.~\ref{eq:source} and \ref{eq:source-rms}, may be derived either for all scans at once; or maps can be produced for each scan first, then coadded after various fine-tuning measures. In the second scenario, post-processing can take place both on individual scan maps and on the composite map, forming a more practical approach to be followed.

\subsection{Excess Noise and Scan Weighting}

While the map noise estimates of Eq.~\ref{eq:source-rms} from the time-stream data are statistically sound, these nonetheless rely on the implicit assumption that both the time-stream noise and the intrinsic map noise are white. The assumption does not always hold true, however: time-streams often come with spectral shapes and can undergo spectral filtering (either explicitly or as a result of decorrelation); maps may also carry $1/f$ type spatial noise as an imprint of sky-noise. If so, the map-pixel noise estimates of Eq.~\ref{eq:source-rms} may become off-target.

Fortunately, as long as the discrepancy is solely attributed to the non-white nature of time-streams and of the source map, the noise difference will spread homogeneously over the entire map. Thus, the ``true'' map noise will differ from the statistical time-stream estimate of Eq.~\ref{eq:source-rms} by a single scalar alone. One can simply measure a representative reduced chi-squared ($\chi_r^2$) from the map itself, and scale the map noise accordingly by $\chi_r$. I.e.,

\begin{equation}
\hat{\sigma}_{xy, {\rm true}} = \chi_r \, \hat{\sigma}_{xy, {\rm calc}}.
\end{equation}

Here, $\chi_r^2$ value may be calculated the usual way $\chi_r^2 = \sum_{xy} w_{xy} \hat{S}^2_{xy} / (N_{xy}-1)$ using noise weights ($w = 1/\sigma^2$) over all map pixels, or over a selected area (e.g.~outside the source). Alternatively, it may be estimated using robust methods (such as described in \ref{sec:estimators}).

\subsection{Direct Line-of-Sight Opacities}
\label{sec:opacities}

The total power level of bolometers offers a measure the optical loading they are exposed to, provided the instrument is sufficiently characterized. For SHARC-2, one can get accurate line-of sight opacities this way. Also, the correlated atmospheric foreground is but variations in optical depth in the line of sight. Therefore, the corresponding correlated component model can act as a real-time measure of the line-of-sight opacity variations, even if the mean value of the optical depth is determined otherwise (e.g.~from skydips or radiometers). Details on how this may be done are offered by Ref.~\citenum{thesis}.

\subsection{Post-processing Steps}

Typical post-processing of maps can include median zero-leveling (typically in the first map generation only), discarding underexposed or excessively noisy map pixels, smoothing and filtering of the larger scales. Time-stream samples containing bright source signals can be identified, and blanked (i.e.~flagged except for source modeling purposes), eliminating or reducing the filtering effect (inverted lobes around sources) for the bright map features with consequent iterations.

Maps produced in the intermediate steps of the analysis can be modified to eliminate unphysical characteristics. For example, when sources are expected to be seen in emission only, the negative features may be explicitly removed, in the hope of aiding convergence towards more physical solutions. However, such tinkering should be avoided when producing a final unbiased source map.

\subsubsection{Smoothing (Convolution by a Beam)}

Owing to the relatively fine gridding recommended for the nearest pixel method, the raw maps will contain noise below the half-beam Nyquist sampling scales of the telescope response. Smoothing can be enlisted to get rid of the unphysical scales, and to greatly improve the visual appearance of images. The smoothing of the source map $\hat{S}(x, y)$ is performed via a weighted convolution by some beam $B(x, y)$, where

\begin{equation}
\hat{S}'(x,y) = \frac { \sum_{u,v} w(u,v) B(u-x,v-y) \hat{S}(u,v) }
{ \sum_{u,v} w(u,v) \left| B(u-x,v-y) \right| },
\end{equation}
\begin{center}
and
\end{center}
\begin{equation}
\hat{\sigma}^2_{S'}(x,y) = \frac { 1 } { \sum_{u,v} w(u,v) \left| B(u-x,v-y) \right| }.
\end{equation}

The convolution is normalized s.t. it preserves integrated fluxes in apertures. Smoothing also increases the effective image beam area by the area of the smoothing beam.

Apart from improving visual appearance, the pixel fluxes and rms values of a beam-smoothed image can be readily interpreted as the amplitudes of fitted beams, and their uncertainties, at the specific map positions\cite{thesis}, and can therefore be used for point sources extraction algorithms.

\subsubsection{Filtering of Large Scales and Wavelet Filters}
\label{sec:filtering}

Filtering of the larger scales, when these are not critical to the astronomer, can often improve the detection significance of compact objects, mainly because maps, like time-streams, tend to have $1/f$ type noise properties. Filtering can be performed by a convolution filter, which first calculates an instance of the map smoothed to the filter scale, then removes this from the original map. One should note that such filtering will reduce the source fluxes by approximately $(D_S/D_{\rm filter})^2$ for characteristic source scales of $D_S$ and a filter FWHM of $D_{\rm filter}$. The loss of source flux can be readily compensated by an appropriate rescaling of the filtered map. The author's implementations adjusts for this for point sources or a specified source size.

Note also, that the combination of Gaussian smoothing and filtering effectively constitutes a wavelet filter, which responds to scales between the smoothing size and the large-scale-structure filtering scale.

\section{DISCUSSION}
\label{sec:discussion}

By now it ought be clear that the inherent complexities of bolometer time-streams require a reduction approach that can simultaneously determine the source, the correlated components, their corresponding detector channel gains, and the correct noise weights. Additionally, one may want to include explicit filtering steps (such as noise whitening or Wiener-filtering), and flag unreliable data.

While the usual matrix methods can solve for all signals at once (assuming gains, data weights, and flags), all the other steps have to be performed separately. Therefore, arriving at a self-consistent solution set of signals, gains, weights, flags and filters, invariably involves an iterated scheme of some sort. Each iteration will entail some or all of the steps: (a) source model estimation (map-making), (b) estimating correlated signals, (c) gain estimation, (d) calculation of noise weights, (e) flagging, (f) explicit filtering.

\subsection{Ordering}

The order, in which reduction steps (i.e., the estimation of correlated signals, gains weights, flagging etc.) are performed can be important. Because {\em CRUSH} splits the estimation of signals (correlated components and source) into individual steps, it provides greater flexibility in arranging these than matrix methods would allow. This has the obvious advantage that the refinement of gains, weights and flags, can proceed as soon as the brightest signals are modeled, benefiting the accuracy of all successive signal estimation steps within the same iteration. In contrast to this, when signals are solved in a single inversion step, the improvement of gains etc.~afterwards can produce results in the next iteration only. For this reason, {\em CRUSH} is expected to converge in faster to self-consistent solutions than matrix methods would.

In light of the above, the right choice of ordering can greatly improve the speed of convergence. Following just a few basic rules can help determine optimal pipeline orderings. Signals, correlated or source, should be ordered such that the brighter signals are solved for first. This way, every step is optimized to leave the cleanest possible residuals behind, hence aiding the accuracy at which successive reduction steps can be performed. Gains should be estimated immediately after the derivation of the corresponding correlated signals. Weighting can be performed as soon as the bright signals, exceeding the white noise level, are modeled, followed flagging of outliers (e.g.~despiking).

The SHARC-2 implementation of {\em CRUSH} can optimize the ordering automatically. First, a quick-and-dirty preliminary iteration (with a default pipeline order) is performed to determine the typical magnitude of component signals. Afterwards, the actual reduction is performed from scratch in order determined by the decreasing fluxes and adherence to the above stated principles.

\subsection{Convergence}

Under optimal pipeline configurations, the convergence of signals, gains weights filters and data flags, can be quickly achieved. The SHARC-2 and LABOCA reductions require just a handful (5--8) iterations, before ``final'' maps are produced. Other instruments may require fewer or more iterations, depending on the complexity of their signal structure. One should expect that the higher the complexity (i.e., the more parameters are to be estimated), the more iterations will be required to arrive at the self-consistent set of solutions.

\subsection{Degeneracies}

Thus far, it has been implicitly assumed that signals can be uniquely separated into the various correlated components and source signals. This is rarely the case, however. Consider, as an example, the case when a bolometer array scans across a very extended source ($D \gg {\rm FoV}$). Invariably, a large part of the source structure is seen by all detectors at once, much like the correlated atmospheric noise is seen by the same pixels. There is no telling apart what part of these correlated signals, seen by all pixels, is source and what part is sky. This presents a dilemma as to how to interpret degenerate signals.

\begin{figure}
\centering
\includegraphics[width=\textwidth]{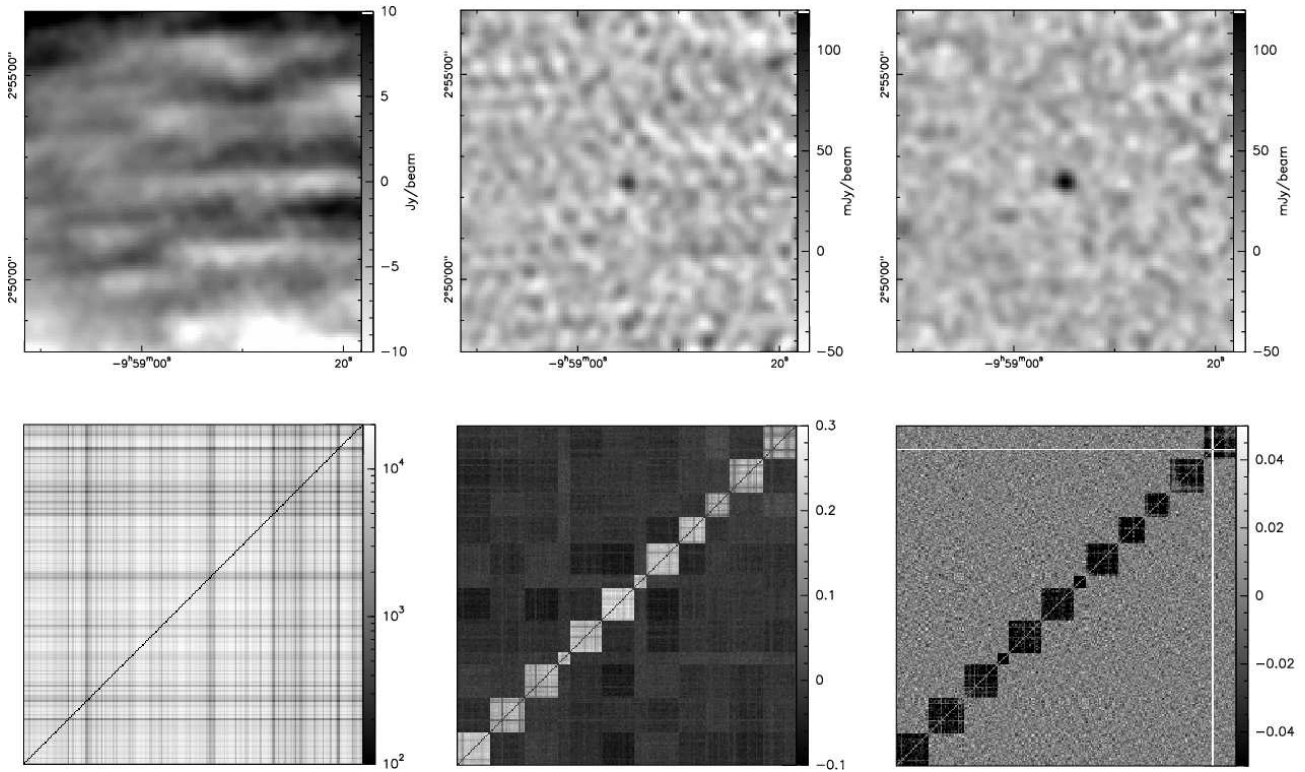}
\vskip 12 pt
\caption{Dealing with degeneracies. The awkward choice between keeping more extended emission or paying the price of higher map noise: an example of a simulated 100\,mJy point source implanted in a single 8-minute blank-field LABOCA scan and reduced three different ways. Shown are a direct map (top left), produced with signal centering only, a map with correlated sky removal (top center), and with additional band-cable decorrelation (top right) taking place before the mapping step. The corresponding effective map rms values are 4.4, 0.012, and 0.011\,Jy/beam respectively. Below the maps are the normalized (see Sec.~\ref{sec:covariances}) residual pixel-to-pixel covariances after the reduction, for the 234 working channels in the array, here with the diagonal 1 values zeroed. The left map preserves source structures on all scales, but these would only be seen if are well in excess of the whopping $\sim$4\,Jy/beam apparent noise level. As the covariance matrix below it demonstrates the data has strong correlated signals across the full array (consistent with atmospheric noise), at levels thousands of times above the detector white noise level. Note, that the larger scales are more severely affected in the map. After removal of the atmospheric noise, the image (top center) no longer contains scales $>$FoV ($\sim$11'), but the noise level drops over two orders of magnitude and the faint inserted source becomes visible. However, the noise is clearly structured and the block-diagonal patterns seen in the covariance matrix below reveal a significant (20--30\% over white noise) correlations within each of the 12 flexible band cables. When these are also modeled prior to the map-making step, one is rewarded with an even cleaner image. At this point, the covariances outside of the decorrelated cable blocks (bottom right) reveal no more correlated signals down to a few percent of the detector white noise levels. However, with the decorrelation of the cables go the scales above the typical footprint of detectors sharing a cable (i.e. $>$0.3--0.5\,FoV). The missing row and column in the covariance matrix is due to a flagged channel in that reduction. The negative covariances left behind by the estimation of correlated cable signals is a visual reminder of the degrees of freedom lost in the modeling step. }
\label{fig:comparison}
\end{figure}

In {\em CRUSH}, owing to the sequential nature of signal estimation, the degenerate signals are normally modeled by the reduction step which estimates these first\footnote{The ``interpretation'' of degenerate signals as source or noise can evolve with iterations but only when noise models are incompletely filtered, or limited in time-resolution (see Ref.~\citenum{thesis} for details).}. Taking the above example of extended source and sky, the degenerate flux ends up in the map if the mapping step precedes the decorrelation across the array. Otherwise, it will form part of the atmospheric noise model. In the first case, the map will contain the extended structure albeit buried in noise, whereas in the latter case one gets a clean looking map but one, which contains no extended emission on scales larger than the field-of-view (Fig.~\ref{fig:comparison}).

In matrix approaches, where the inversion step solves for correlated components and source flux distribution simultaneously, the destination of degenerate signals is often nontrivial and obscured. For direct inversion methods, the degeneracies manifest as singularities that make direct inversion impossible until these are explicitly decoupled by appropriate constraining. However, as the degeneracies can be multiple and their relations often complex, this is not a very practical route to take.

Instead, the more common approach is to use pseudo-inversion methods, like Singular-Value Decomposition (SVD), which always produce a $\chi^2$-minimizing solution. The problem is that SVD picks just one such solution from a possible family of equally good solutions, and that this pick is controlled by a mathematical contraint rather than a physical one. Back to our example of degenerate source and sky, interpreting a part $\alpha$ of the degenerate signals as source flux and the remaining $(1-\alpha)$ part as correlated atmosphere, represents formally an equally good solution (i.e.~with identical $\chi^2$) for all values of $\alpha$. However, SVD effectively chooses for us some value of $\alpha$, based on a purely mathematical idea. One has no effective control over what that value will be. Thus, maps produced by SVD can have arbitrary amounts of the degenerate noise inserted in their source maps. Worse, one cannot easily know how much that really is.

The explicit control {\em CRUSH} offers, simply by the pipeline ordering, over the interpretation of degenerate signals, is then perhaps the greatest conceptual advantage of {\em CRUSH} over matrix methods.

\subsection{Limitations to the Recovery of Extended Emission}

As seen above, the degeneracies between source and correlated signals present the astronomer with an unattractive choice between producing a more complete source map albeit a noisy one, or producing a cleaner map but one which lacks structural components (e.g.~extended emission $>$FoV, see Fig.~\ref{fig:comparison}). The awkward nature of this choice can be mitigated only by better observing strategies that render such degeneracies less in the way of scientific results. Whenever maximal sensitivity (i.e., minimal noise) is required, one has no real choice but to put up with incomplete (i.e.,~filtered) models of the source.

In case of astronomical imaging, the problem of degeneracies typically manifests itself on the larger spatial scales. Source emission on scales greater than the smallest typical footprint (on sky) of a group of detectors with the same correlated signal component will be difficult (if not outright impossible) to recover.

The filtering effect of decorrelating steps is easily characterized for the case of maximum-likelihood estimators. Consider Eq.~\ref{eq:estimator} for the estimation of a correlated component. If the astronomical source flux of interest is simultaneously expected in {\bf E}$(N_S)$ detectors alongside the correlated signals present in a group of channels, then a fraction

\begin{equation}
\phi_c \approx (1-\Phi)\, {\rm \bf E}(N_S)\, \frac { w_c g_c^2} { \sum_c w_c g_c^2}
\label{eq:filtering}
\end{equation}

of the source flux in the time-stream of channel $c$ will fall casualty to the decorrelation step after the filtering of the model signals (Sec.~\ref{sec:model-filtering}). One can recompense by appropriately including a factor $(1-\phi_c)$ in the effective source gain (see Sec.~\ref{sec:source}). For close-packed feedhorn arrays, like LABOCA, {\bf E}$(N_S) = 1$ for source sizes up to the two beam spacing of the horns. For filled arrays {\bf E}$(N_S) \approx A_S/A_{\rm det}$ in terms of the effective source area $A_S$ and detector pixel area $A_{\rm det}$. Naturally, such corrections apply only to one source scale at a time. When a range of source scales are observed, the best one can do (within this compensation scheme) is to apply the corrections for the median scale and hope that the residual $\delta \phi_c$ are typically small enough to be absorbed within the typical calibration uncertainties for the other scales. Such corrections form part of the author's software packages, but not of the other implementations at present.

\subsection{Reducing Scans Together vs.~Individually}

Some of the harmful degeneracies may be dependent on orientation or the particular scanning configuration\cite{scanning} and, thus, can change from scan-to-scan. This may move these degeneracies into different source components. The correlated detector rows of SHARC-2 are an example if this. In a given orientation $x$ they are degenerate with all the spectral source components $\tilde{S}(\omega_x, 0)$. By rotating the array relative to the mapped field, these degeneracies will manifest along a different direction in the source map. In practice, the field rotation can be realized without explicit instrument rotation, simply by letting the source transit across the sky.

What this means from the reduction point of view is that the degeneracies, like the correlated detector rows, will affect single scans with negligible field rotation whereas a set of scans spanning more rotation of the sky will be immune from such an oriented degenerate condition. It turns out that {\em CRUSH} naturally recovers those source components when models of the source (derived from the full set of scans) and the multioriented degenerate components are iterated\cite{thesis}. Therefore, it is always advisable to reduce large sets of scans together to avoid unnecessary filtering by the decorrelation steps.

The bundled reductions, with their composite source models, are also better able to discriminate problems that may be specific to single scans, simply by increasing the redundancy of data under joint analysis. One should always reduce entire data sets, or as many scans at once as is feasible.

\subsection{Parallelization and Distributed Computing}

Because most of the reduction steps treat scans (or equivalent blocks of data sharing the exact set of channel gains and weights) independently, these can be computed on separate CPUs or networked computing nodes. Only the source map needs to be computed from the full data set. Therefore, the exchanging the source model among the otherwise independent reduction processes opens up the possibility of reducing extremely large data sets on a networked cluster of affordable computers.

\subsection{Pre-processing steps}

Before data is crunched through the pipeline analysis of {\em CRUSH} (or other methods), one should make full use of all the external information available that does not form integral part of the reduction itself. Data that are clearly not useful or problematic (e.g.~dead or cross-talking channels or frames with too high telescope accelerations that could induce mechanical power loads) should be discarded immediately. Frames with unsuitable mapping speeds for use in the map-making step should be flagged accordingly. Another example would be correcting for cold-plate temperature fluctuations using either thermistor data, or the readout from blinded bolometers if and when available. Any other measures that improve data quality should be taken prior to the estimations of parameters by the reduction.

\subsection{CRUSH vs.~Principal Component Analysis (PCA)}

Principal Component Analysis\cite{pca} (PCA) is a powerful alternative method. By diagonalizing (measured) pixel-to-pixel covariance matrices (cov$_{ij} \sim \left<(X_{it} - \mu_i) (X_{jt} - \mu_j)\right>_t$), it can ``discover'' the dominant correlated signals, i.e. the eigenvectors with the largest variances (eigenvalues), which characterize the data. (Note, that the estimation of the covariance matrix effectively produces noise weights and gains for all correlated components.) The first $k$ most dominant components can be duly removed from the time-streams. While this seems attractive, the method has some important flaws.

First, the choice of $k$ correlated components is somewhat arbitrary. The removal of an insufficient number or otherwise too many components will result in under- or overfiltering of data when compared to targeted methods (e.g.~{\em CRUSH}). A further complication is that pure source vectors may, at times, slip among the dominant set of $k$ vectors. When they do, such source components are wiped out unnecessarily; at the same time, more of the noise will survive. For these reasons, the source filtering properties of PCA are both obscure and unpredictable, while its noise rejection is often sub-optimal.

Besides, while the PCA method itself is linear with data volume, the estimation of the pixel-to-pixel covariances is an $\mathcal{O}(N^2_{\rm pix}\times N_t)$ process with the pixel count $N_{\rm pix}$ and $N_t$ frames.  This alone could render such methods unfeasible for the ever growing imaging arrays of the future. Neither is the estimate of the covariance matrix from the data necessarily representative of the underlying correlations. Finite data sets tend to underestimate $1/f$ type noise covariances. As such noise commonly affects bolometer signals, this may ultimately limit the usefulness of PCA in these applications.

In conclusion, PCA is best used for exploratory analysis only. Targeted methods, like {\em CRUSH} or SVD, should always take over once the nature of correlations is understood.

\subsection{Learning from the Data}
\label{sec:covariances}

Calculating pixel-to-pixel covariance matrices, especially in normalized form (i.e., $\hat{K}_{ij} = {\rm cov}_{ij} / {\sigma_i \sigma_j}$), similar to what PCA would use (above), can be tremendously useful for identifying what correlated components may be present in the time-stream data of an array instrument (Fig.~\ref{fig:comparison}). This may be one of the most powerful tools at hand that can create understanding of the physical characteristics of instruments, and help decide what targeted signal models one should use for optimal results.

\section{CONCLUSIONS}

The data reduction approach pioneered by {\em CRUSH} presents a formidable alternative to matrix approaches in providing a capable and fast data reduction scheme for future large imaging bolometer arrays. The advantages of the approach are best harvested for ground-based instruments which operate under a dominant and variable atmospheric foreground, or with other sources of correlated interference. The approach is conceptually simple and allows fine-tuning capabilities that matrix methods cannot offer. It also deals with the inherent degeneracies of data more transparently than matrix methods do, possibly leading to superior quality images at the end of the reduction.

Computational considerations make the {\em CRUSH} approach especially attractive. Its computing requirements (both memory and number of operations necessary) scale strictly linearly with increasing data volume. Moreover, the method is well suited for cluster computing, allowing faster reductions of data sets, and the coherent reduction of data sets many times larger than single machine RAM capacity. For these reasons, {\em CRUSH} may yet come to represent the best option for providing data reduction capabilities to a whole range of future instruments with large data volumes.

Neither is the approach necessarily limited to imaging bolometers alone. The same data reduction philosophy has direct application for all instruments that operate a number of channels under correlated noise interference, especially~in scanning observing modes (i.e.~where the source signals are moved from channel-to-channel). For example, channels may be frequency channels of a spectroscopic backend, and scanning may take place in frequency space. Adaptations may become possible for interferometric applications also (e.g.~ALMA), but this idea remains to be investigated.

\acknowledgements

The author wishes to thank Darren Dowell, who contributed many excellent ideas to {\em CRUSH}, and who has been the most steadfast tester and bug-reporter throughout the years. Tom Phillips deserves gratitude for having supported the long process of developing this new approach under the umbrella of PhD research funded by the NSF. Last but not least, many thanks to the numerous patient users of {\em CRUSH} who have endured countless software bugs, but haven't (yet) given up on it entirely.


\end{document}